\documentclass{PoS}
\usepackage{graphicx}

\input psfig.sty
\def \trho{{\tilde{\rho}}}

\title{Accretion disks with a large scale magnetic field around black holes, and magnetic jet collimation}

\ShortTitle{Magnetized accretion disks and jets}

\author{\speaker{G.S. Bisnovatyi-Kogan}\\
        Space Research Institute, Russian Academy of Sciences, Moscow, Russia\\
        E-mail: \email{gkogan@iki.rssi.ru}}

\author{ R.V.E.~Lovelace\\
        Cornell University, Ithaca, USA\\
        E-mail: \email{lovelace@astro.cornell.edu}}

\abstract{ We discuss the problem of the formation of a large-scale
magnetic field in the accretion disks around black holes, taking
into account the nonuniform vertical structure of the disk. The high
electrical conductivity of  the outer layers of the disk prevents
the outward diffusion of the magnetic field.
  This implies a stationary state with a strong magnetic field in
the inner parts of the accretion disk close to the black hole, and
zero radial velocity at the surface of the disk. Magnetic jet collimation
is considered, when the jet radius is hold due to magneto-torsional oscillations.
The range of parameters is found where jet radius is oscillating,
regularly or chaotically, within restricted values.  }

\FullConference{25th Texas Symposium on Relativistic Astrophysics - TEXAS 2010\\
        December 06-10, 2010\\
        Heidelberg, Germany}

\begin{document}

\section{Introduction}
Quasars and AGN contain supermassive black holes, about 10 HMXR
contain stellar mass black holes - microquasars. Jets are observed
in objects with black holes: collimated ejection from accretion
disks.

   Early work on disk accretion to a black hole
argued that a large-scale magnetic field of, for example, the
interstellar medium would be dragged inward and greatly compressed
by the accreting plasma \cite{2,3,5}.
 Subsequently, analytic models of the
field advection and diffusion in a turbulent disk suggested, that
the large-scale field diffuses outward rapidly \cite{11,7}, and
prevents a significant amplification of the external poloidal field
by electrical current in the accretion disk.
   This has led to the suggestion that special conditions
(non-axisymmetry) are required for the field to be advected inward
\cite{15}.
    The question of the advection/diffusion
of a large-scale magnetic field in a turbulent plasma accretion disk
was reconsidered in \cite{bkl07}, taking into account its nonuniform
vertical structure.
   The high electrical conductivity of the surface layers
of the disk, where the turbulence is suppressed
 by the radiation flux and the relatively high magnetic field,  prevents
outward diffusion of the magnetic field. This leads to  a strong
magnetic field in
 the inner parts of accretion disks around black holes.

\section{The fully turbulent model}

  There are two limiting accretion disk models which have analytic
solutions for a large-scale magnetic field structure.
   The first was constructed in \cite{2,3} for
a stationary non-rotating accretion disk.
   A stationary state in
this disk (with a constant mass flux onto a black hole) is
maintained by the balance between magnetic and gravitational forces,
and thermal balance (local) is maintained by Ohmic heating and
radiative conductivity for an optically thick conditions.
   The mass flux to the black hole in the accretion disk is determined by
the finite conductivity of the disk matter and the diffusion of
matter across the large-scale magnetic field as sketched in Fig.1.
 The value of the large-scale magnetic
field in stationary conditions is determined by the accretion disk
mass, which in turn is determined by the magnetic diffusivity of the
matter.
\begin{figure}
\centerline { \psfig{figure=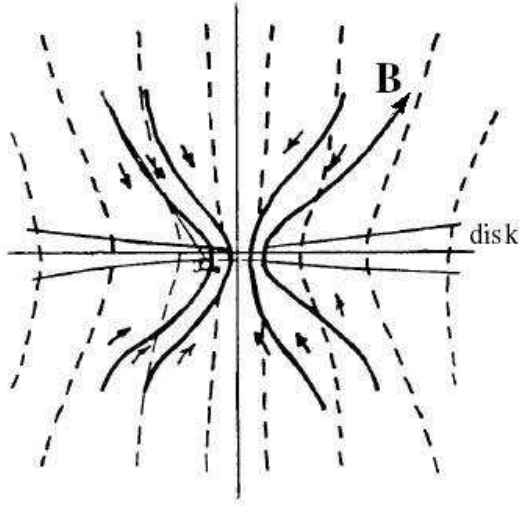,width=8cm}} \caption{Sketch of
the poloidal magnetic field threading an accretion disk. The field
strength increases with decreasing radius
 owing to flux freezing in the accreting
disk matter, from \cite{3}.}
  \label{fig4}
\end{figure}
  It is widely accepted that the laminar disk is unstable to
different hydrodynamic, magnetohydrodynamic, and plasma
instabilities which implies that the disk is turbulent.
  In X-ray binary systems the assumption about turbulent accretion
disk is necessary for construction of a realistic models \cite{ss}.
The turbulent accretion disks had been constructed also for
non-rotating models with a large-scale magnetic field.
   A formula for turbulent magnetic diffusivity was
derived in \cite{3}.
  similar to the scaling of the shear $\alpha$-viscosity
in turbulent accretion disk in binaries \cite{ss}, where the viscous
stress tensor component $t_{r\phi}=\alpha P$, with $\alpha \leq 1$ a
dimensionless constant and $P$ the pressure in the disk midplane.
disk.
  Using this
representation, the expression for the turbulent electrical
conductivity $\sigma_t$ is written as
\begin{equation}
 \label{eq1}
 \sigma_t=\frac{c^2}{\tilde\alpha 4 \pi h \sqrt{P/\rho}}.
\end{equation}
Here, $\tilde\alpha = \alpha_1 \alpha_2$.
  The characteristic
turbulence scale is $\ell=\alpha_1 h$, where $h$ is the
half-thickness of the disk, the characteristic turbulent velocity is
$v_t = \alpha_2 \sqrt{P/\rho}$.
   The  large-scale magnetic field threading
a turbulent Keplerian disk arises from  external electrical currents
and currents in the accretion disk.
   The field generated by the currents in
the disk can be much
larger than that due to the external currents.
   The magnetic field may become
dynamically important, influencing the accretion disk structure and
leading to powerful jet formation, if it is strongly amplified
during the radial inflow of the disk matter. It is possible only
when the radial accretion speed of matter in the disk is larger than
the outward diffusion speed of the poloidal magnetic field due to
the turbulent diffusivity $\eta_t=c^2/(4\pi \sigma_t)$.
  Estimates in \cite{11} have shown
that for a turbulent conductivity (\ref{eq1}),
the outward diffusion
speed is larger than the accretion speed.
  Thus it appears that there is no large-scale
magnetic field amplification during Keplerian disk accretion.
Numerical calculations in \cite{11} are reproduced analytically for
the standard accretion disk structure which can be written as
(e.g.\cite{3a})
\begin{equation}
\label{eq2}
 \dot M = 4\pi \rho v_r rh~,\,\, h=\frac{v_s}{\Omega_K}~,\,\,
 v_s=\sqrt\frac{P}{\rho}~, \,\, 4\pi r^2 h\alpha P=\dot M(j-j_{in})~,
\frac{3}{2}\frac{\Omega_K}{r},\,\, \alpha Prh=\frac{2aT^4c}{3\kappa
\rho h}~.
\end{equation}
 Far from the inner disk boundary the specific angular
momentum is $j \gg j_{in}$. The characteristic time $t_{visc}$ of
the matter advection  due to the shear viscosity is $
t_{visc}=\frac{r}{v_r}=\frac{j}{\alpha v_s^2}~$. The time of the
magnetic field diffusion is $t_{diff}=\frac{r^2}{\eta}\frac{h}{r}
\frac{B_z}{B_r}~,\,\,\,
 \eta=\frac{c^2}{4\pi\sigma_t}=\tilde\alpha h v_s$.
In the stationary state, the large-scale magnetic field in the accretion
disk is determined by the equality $t_{vis}=t_{diff}$, what
determines the ratio
\begin{equation}
\label{eq6} \frac{B_r}{B_z}=\frac{\alpha}{\tilde\alpha}
\frac{v_s}{v_K}= \frac{\alpha}{\tilde\alpha} \frac{h}{r}\ll 1~.
\end{equation}
Here, $v_K=r \Omega_K$ and $j=rv_K$ for a Keplerian disk.
   In a fully
turbulent disk a matter is penetrating through magnetic field lines,
almost without a field amplification. Note, that the field induced
by the azimuthal disk currents has $B_{zd} \sim B_{rd}$ \cite{21}.

\section{Turbulent disk with radiative outer zones}

  Near the surface of the disk, in the
region of low optical depth, the turbulent
motion is suppressed by the radiative flux, similar to the suppression of the
convection over the photospheres of stars with outer convective zones.
  The presence of the outer radiative layer does not affect the
estimate of the characteristic time $t_{visc}$ of the matter
advection in the accretion disk because it is determined by the
main turbulent part of the disk.
  The time of the field diffusion, on the
contrary, is significantly changed, because the electrical current is
concentrated in the radiative highly conductive regions, which
generate the main part of the magnetic field.
  The structure of the
magnetic field with outer radiative layers is shown schematically in
Fig.2.
\begin{figure}
\centerline{\psfig{figure=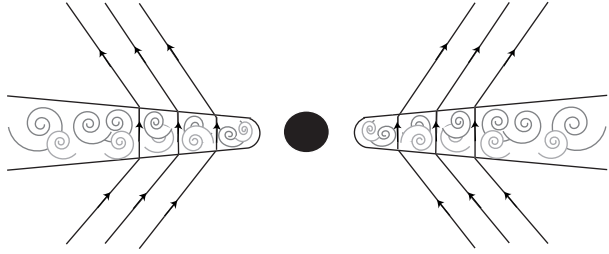,width=8cm}} \caption{Sketch of the
large-scale poloidal magnetic field threading a rotating turbulent
accretion disk with a radiative outer boundary layer.
  The toroidal current flows mainly in the highly
conductive radiative layers.
  The large-scale (average)
field in the turbulent region is almost vertical.}
 \label{fig5}
\end{figure}
 Inside the turbulent disk the electrical current is
negligibly small so that the magnetic field there is almost fully
vertical, with $B_r \ll B_z$, according to (\ref{eq6}).
    In the outer radiative layer, the field
diffusion is very small, so that matter advection is leading to
strong magnetic field amplification. We suppose, that in the
stationary state the magnetic forces could support the optically
thin regions against gravity.
  When the magnetic force balances the gravitational force in the
outer optically thin part of the disk of surface density
$\Sigma_{ph}$ one finds the following relation takes place \cite{3}
\begin{equation}
\label{eq8} \frac{GM\Sigma_{ph}}{r^2}\simeq \frac{B_z
I_\phi}{2c}\simeq \frac{B_z^2}{4\pi}~,
\end{equation}
 The surface density over the photosphere corresponds to a layer
with effective optical depth close to $2/3$ (e.g. \cite{21}).
  We estimate the lower limit of the magnetic field strength,
taking $\kappa_{es}$ (instead of the effective opacity
$\kappa_{eff}=\sqrt{\kappa_{es}\kappa_a}$). Writing
$\kappa_{es}\Sigma_{ph}=2/3~$, we obtain $\Sigma_{ph}=5/3$
(g/cm$^2)$ for the opacity of the Thomson scattering,
$\kappa_{es}=0.4$ cm$^2$/g.
  The absorption opacity $\kappa_a$ is
much less than $\kappa_{es}$ in the inner regions of a luminous
accretion disk.  Thus using in equation (\ref{eq8}), the above
$\Sigma_{ph}$, we estimate the lower bound on the large-scale
magnetic field of a Keplerian accretion disk as
\begin{equation}
\label{eq10}
B_z=\sqrt{\frac{5\pi}{3}}\frac{c^2}{\sqrt{GM_\odot}}\frac{1}{x\sqrt{m}}\simeq
10^8{\rm G} \frac{1}{x\sqrt{m}}, \quad x=\frac{r}{r_g}, \quad
m=\frac{M}{M_\odot}~.
\end{equation}
The maximum magnetic field is reached when the outward magnetic force
balances the gravitational force on the disk of surface mass
density $\Sigma_{ph}$.
In equilibrium, $B_z \sim \sqrt{ \Sigma_{ph}}$.
   We find that $B_z$ in a
Keplerian accretion disk is about $20$ times less than its maximum
possible value, for $x=10,\,\,\alpha=0.1,$ and  $\dot m=10$.

\section{Self-consistent numerical model}

   Self-consistent models of the rotating
accretion disks with a large-scale magnetic field requires solution
the equations of magnetohydrodynamics.
 In presence of the radiative layer the strength of the magnetic field is large,
and it  may greatly  exceed the strength of the seed field.
 The solution with a small field will not be stationary,
 and a transition to the strong field solution will take place.
Therefore the strong field solution is the only stable stationary
solution for a rotating accretion disk. The vertical structure of
the disk with a large scale poloidal magnetic field was calculated
in \cite{lrb}, taking into account the turbulent viscosity and
diffusivity, and the fact that the turbulence vanishes at the
surface of the disk. The full system of equations was reduced to one
vertical equation for the non-dimensional radial velocity $u_r$, in
the form
$$
{\alpha^4\beta^2} {\partial^2 \over \partial \zeta^2}
\left(g{\partial \over \partial \zeta} \left(\trho g{\partial \over
\partial \zeta} \left({1\over \trho}{\partial \over \partial \zeta}
\left(\trho g {\partial u_r \over \partial \zeta}\right)\right)
\right)\right)
$$$$
-~\alpha^2\beta {\cal P} {\partial^2 \over \partial \zeta^2} \left(g
{\partial \over \partial \zeta} \left(\trho g{\partial \over
\partial \zeta} \left({u_r \over \trho g}\right)\right)\right)
-~\alpha^2\beta {\cal P} {\partial^2 \over \partial
\zeta^2}\left({1\over \trho} {\partial \over \partial \zeta}
\left(\trho g {\partial u_r \over \partial \zeta}\right)\right)
$$
\begin{equation}
+~{\alpha^2\beta^2 } {\partial^2 \over \partial \zeta^2}\bigg(\trho
g \big(u_r- gu_0 \big)\bigg) +{\cal P}^2 {\partial^2 \over \partial
\zeta^2}\bigg({u_r \over \trho g} \bigg)
\nonumber \\
+~3\beta{\cal P }^2{u_r \over g}=0~.
\end{equation}
Here  $\zeta \equiv z/h$ is a dimensionless height, $u_r \equiv
-~v_r/(\alpha c_{s0})$, $u_0$ is a non-dimensional radial velocity
in the non-magnetized disk \cite{ss}. Coefficients of the turbulent
viscosity $\nu$, and magnetic diffusivity $\eta$ are connected by
the magnetic Prandtl number $\cal P \sim$1, $ \nu ={\cal P} \eta
=\alpha ~{c_{s0}^2 \over \Omega_K}~ g(z)~, $ where $\alpha$ is a
constant, determining the turbulent viscosity \cite{ss}; $\beta
=c_{s0}^2/v_{A0}^2$, where $v_{A0}=B_0/(4\pi\rho_0)^{1/2}$ is the
midplane Alfv\'en velocity, $\tilde{\rho}={\rho \over \rho_0}$. The
function  $g(z)$ accounts for the absence of turbulence in the
surface layer of the disk \cite{bkl07}.
    In the  body of the disk $g = 1$, whereas
at the surface of the disk, at say $z_S$, $g$ tends over a short
distance to a very small value, effectively zero. The smooth
function with a similar behavior is taken \cite{lrb} in the form
 $
g(\zeta)=\left(1-{\zeta^2\over \zeta_S^2}\right)^\delta ~,
 $
with $\delta \ll 1$. In the stationary state the boundary condition
on the disk surface is  $u_r=0$, and only one free parameter -
magnetic Prandtl  number $\cal P$ remains in the problem. In a
stationary disk vertical magnetic field has a unique value. The
example of the radial velocity distribution for $\cal P=$1 is shown
in Fig.3.
\begin{figure}
\centerline{\psfig{figure=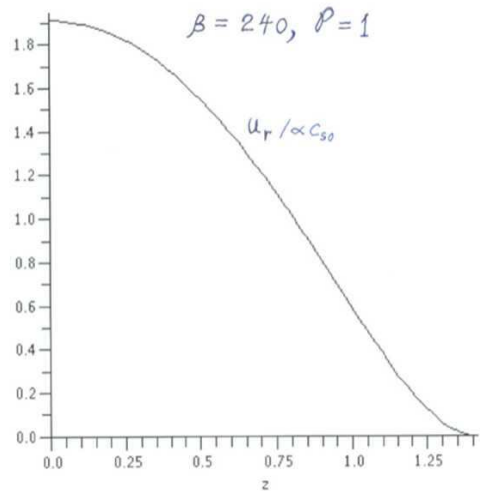,width=11cm}}
\caption{Distribution of the radial velocity over the thickness in
the stationary accretion disk with a large scale poloidal magnetic
field}
 \label{fig6}
\end{figure}

\section{Jet collimation}

Magnetic collimation is connected with torsional oscillations of a
cylinder  with elongated magnetic field, see Fig.4. The stabilizing
azimuthal magnetic field is created by torsional oscillations.
Approximate simplified model is developed \cite{mn07}. Ordinary
differential equation is derived, and solved numerically, what gives
a possibility to estimate quantitatively the range of parameters
where jets may be stabilized by torsional oscillations.

\begin{figure}
\centerline{\psfig{figure=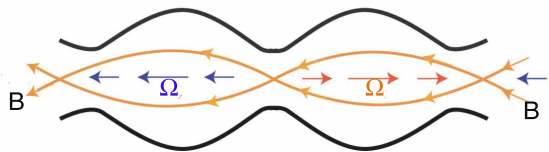,width=8cm}} \caption{Jet
confinement by magneto-torsional oscillations (qualitative picture)}
\label{fig7}
\end{figure}

In non-dimensional variables $\tau=\omega t, \,\, y=\frac{\tilde
R}{R_0},\,\, z=\frac{a \tilde R}{a_0 R_0},\,\,
 a_0=\frac{K}{\omega R_0^2}=\omega,\,\,R_0=\frac{\sqrt K}{\omega}$,
differential equations have a form
\begin{equation}
\label{eq54} \frac{dy}{d\tau}=z,\, \frac{d
z}{d\tau}=\frac{1}{y}(1-D\sin^2 \tau),\, y(0)=1,\, z=0\,\,{\rm
at}\,\, \tau=0.
\end{equation}
The system (\ref{eq54}) has two non-dimensional parameters:
$D=\frac{1}{2\pi K
C_m}\left(\frac{C_b\Omega_0}{z_0\omega}\right)^2$, and $y(0)$, and
the second one is taken equal to unity . Solution  changes
qualitatively with changing of the parameter $D$. At $D \le 2$ there
is no confinement, radius grows to infinity after several
low-amplitude oscillations. At $D=2.1$ radius is not growing to
infinity, but is oscillating around some average value, forming
complicated curves (Fig. 5).At $D \ge 2.28$ the radius goes to zero.
On the edge of the cylinder the rotational velocity cannot exceed
the light velocity. The analysis have shown \cite{mn07} that for the
sound velocity not exceeding $c/2$, the jet should contain baryons,
density $\rho_0$ exceeding about 30\% of
 the total density of the jet.
Development of chaos in this system, by construction of  Poincare sections
was investigated in \cite{bknstk}.

\begin{figure}
\centerline{\hbox{\includegraphics[width=0.5\textwidth]{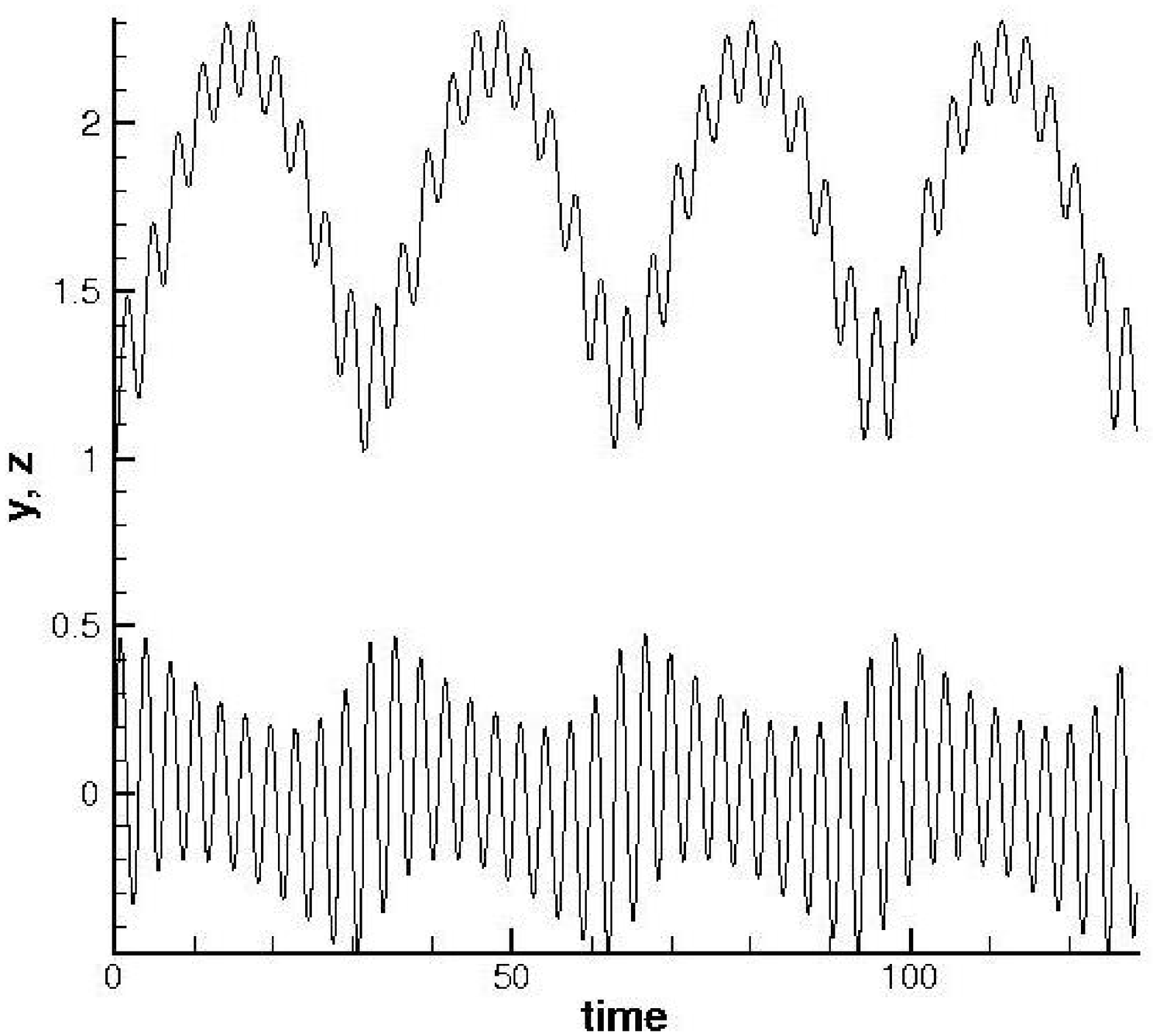}}
            \hbox{\includegraphics[width=0.5\textwidth]{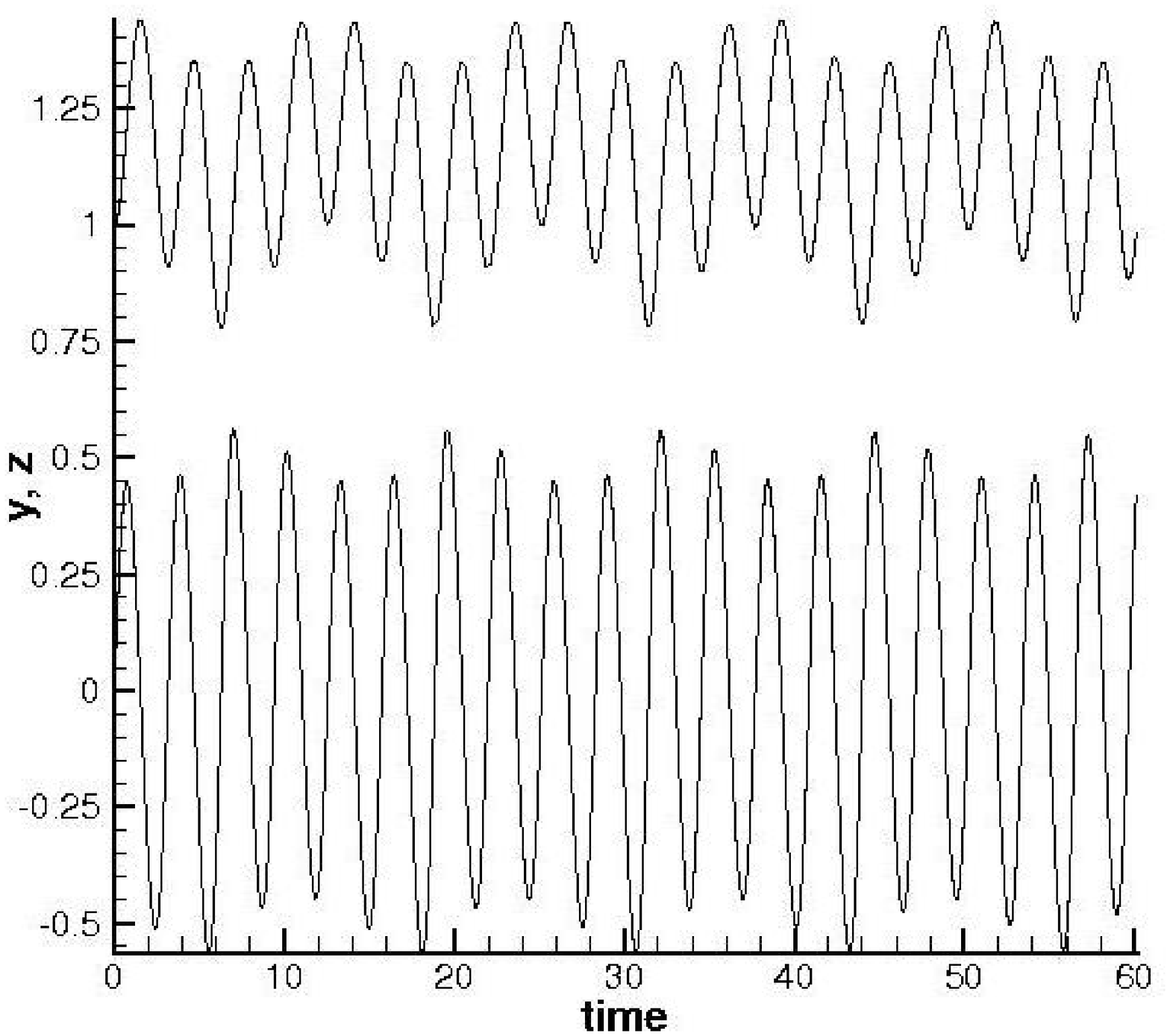}}}
\caption{Time dependence of non-dimensional radius $y$ (upper
curve), and non-dimensional velocity $z$ (lower curve), for $D=2.11$
(left); for $D=2.25$ (right), from \cite{mn07}.}
\end{figure}

\section{Conclusions}

1. Disk field is amplified during disk accretion due to high
conductivity in outer radiative layers. Stationary solution
corresponds  to $\beta=240$ for Pr=1.

2. Jets from accretion disk are magnetically collimated by
large scale poloidal magnetic field by torsion oscillations,
which may be regular or chaotic.

\end{document}